\documentstyle[12pt,epsfig]{article}
\begin{document}

\language=0

\title{CalcHEP 2.3: MSSM, structure functions, event generation, batchs,
and generation of  matrix elements for other packages.} 

\author{A.Pukhov\\ 
Moscow State University, Russia.
      }
\date{}
\maketitle
\abstract{CalcHEP  is a  package for computation  of  Feynman diagrams and 
integration over  multi-particle phase space.  The main idea prescribed into
CalcHEP is to
make available passing on from  Lagrangians to the final distributions
effectively with a high level of automation. This article presents 
new options of CalcHEP available in version 2.3. They are a) MSSM model
with different SUGRA scenarios, b) interface with PDFLIB and implementation 
of new MRST/CTEQ structure functions, c) realization of approach to 
structure functions for models with diagonal CKM, d) generation of events 
and interface with PYTHIA, e) calculations in non-interactive (batch) 
regime, f)generation of code of different matrix elements for other 
programs, g) many new interface facilities.

}         

\section{Introduction}            \label{introduction}

 CalcHEP is a package for automatic calculations of elementary particle
decay and collision properties in the lowest order of perturbation theory
(the tree approximation). CalcHEP is a next development of the CompHEP\cite{COMPHEP} package 
which was created by the author together with  his colleagues in 
Skobeltsyn Institute of Nuclear Physics. Other packages  created to solve  a 
similar problem are  
FeynArt\cite{FeynArt}, GRACE\cite{GRACE}, HELAS\cite{HELAS}, 
MADGRAPH\cite{MADGRAPH}, PHELAS\cite{PHELAS}, O'MEGA\cite{OMEGA}.    
See also the review \cite{Harlander}.

\begin{figure}[ht] 
\begin{picture}(450,300)
\put(0,0){\epsfig{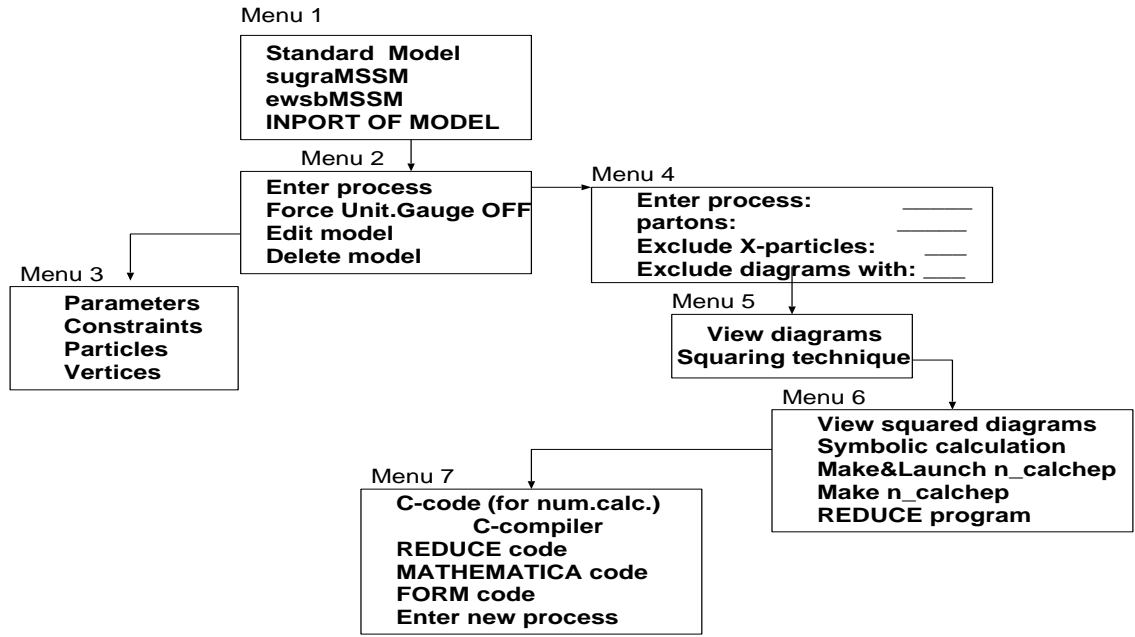}}
\end{picture}

\caption{ Menu scheme for the  symbolic session }

\label{s_chain}

\end{figure}

  CalcHEP is a menu-driven system with the context help. The notations
used in CalcHEP are very similar to those used in particle physics. 
  The CalcHEP package consists of two parts: symbolic and numerical.
The symbolic part produces  C  codes for a squared matrix
element, and they  are used in the numerical calculation  later on.
Menu systems for both parts and relations between  them are shown 
on Fig.\ref{s_chain} and Fig.\ref{n_chain}. 

\vskip 0.2cm  
\noindent
The symbolic part of CalcHEP lets the user:

\begin{itemize}

\item select a model of particle interaction and implement some 
changes in the model. In particular one can choose the package for
solution of RGE equations in case of SUGRA models;

\item choose a gauge between  the physical  or t'Hooft-Feynman ones.

\item  select a process by specifying incoming and outgoing particles for
the decays of $1 \rightarrow 2, \ldots ,1 \rightarrow 5$ types and the collisions of $2
   \rightarrow 2, \ldots , 2 \rightarrow 4$ types;
   
\item   generate  Feynman diagrams,  display them, and create the
corresponding \LaTeX~ output;

\item   exclude some diagrams;

\item   generate and display squared Feynman diagrams;

\item   calculate analytical expressions corresponding to squared diagrams
   by using the fast built-in symbolic calculator; There is an additional 
option  to perform the calculations in the leading order of $1/N_c$
expansion.
   
\item   save symbolic results corresponding to the squared diagrams
   calculated in the REDUCE and MATHEMATICA
codes for  further
   symbolic manipulations;

\item   generate the optimized  C  codes for the squared matrix
   elements for further numerical calculations;

\item   launch the compilation of the generated codes and start the 
corresponding numerical session;
     
\item   generate libraries of matrix elements for other packages.

\end{itemize}

\noindent
The numerical part of CalcHEP offers to:

\begin{itemize}
\item  convolute  the squared matrix element  with structure functions and
beam spectra. The PDFLIB, CTEQ, and MRST   parton
   distribution functions, the  ISR and  Beamstrahlung spectra of electrons,
   the laser photon spectrum,
   and the Weizsaecker-Williams photon structure functions are
   available;

\item   modify physical parameters (incoming momenta, couplings, masses etc.)
   involved in the process;

\item   select the scale parameter for evaluation of the QCD coupling constant and
   parton structure functions;
\item calculate Higgs widths and decay rates taking into account 
high order QCD loop corrections.

\item apply LEP mass limits on MSSM spectrum, as well as 
calculate $b \to s\gamma$, $B_s \to \mu^+ \mu^-$, and $(g-2)_{\mu}$
constraints.

\item  apply various kinematic cuts.

\item   define the kinematic scheme (phase space parameterization) for
   effective Monte Carlo integration;

\item introduce a phase space mapping in order to 
smooth sharp peaks of a squared matrix element and structure functions;

\item   perform  a   Monte Carlo phase space  integration by VEGAS;

\item   generate partonic level  events and direct them to PYTHIA 

\item   display distributions in various kinematic variables;

\item    create the graphical and LaTeX outputs for histograms.

\end{itemize}

The current version is accompanied  with various {\it batch}
programs which allow to performs all calculations in non-interactive 
regime.

CalcHEP source codes for Unix  and  complete manual are disposed 
 on the following Web sites\\
\hspace*{2cm}\verb| http://theory.sinp.msu.ru/~pukhov/calchep.html|\\
\hspace*{2cm}\verb| http://www.ifh.de/~pukhov/calchep.html|\\
In this paper we describe new option of CalcHEP available staring from version 2.3.

\paragraph{A comment about file structure of CalcHEP.}

 We  use notation \verb|$CALCHEP|\footnote{It is up to the user 
responsibility to define the environment variable \verb|CALCHEP|
to simplify his/her work with CalcHEP. In general it is not needed.} 
to designate the  CalcHEP code disposition 
on the disk space. During installation  the user creates another
directory, say \verb|WORK|, intended for user files. \verb|WORK| contains
subdirectories \\
\verb|  models/  tmp/   results/  bin/ |\\
intended for user's version of models, temporary files, numerical sessions 
and CalcHEP commands  respectivitely.  Note that \verb|bin| is just a 
symbolic link of \verb|$CALCHEP/bin/| directory. 

\verb|WORK/| contains 
also a startup script command  \verb|calchep| which  launches symbolic session
\verb|$CALCHEP/bin/s_calchep| which it its turn  realizes work with models, 
symbolic calculations and the record of the obtained matrix elements into 
\verb|WORK/results|. After that \verb|calchep| calls C-compiler and 
transforms C-code stored in \verb|WORK/results| into exequtable 
\verb|n_calchep| disposed in the same place.  The last command 
realises CalcHEP numerical calculations.

\section{New service facilities.}  \label{newservice}

Version 2.3 has some  new service facilities
which simplify  interactive sessions and also 
are necessary  to realize some batch commands. 
\paragraph{ In case of work with 
CalcHEP tables} (model files, cuts, distributions) the following new 
options are implemented:

a) Search a  record in the table, which contains simultaneously  
several pattens. The search is activated by the \verb|^F| key. 
After that the user has to introduce the pattens separated by commas.
 This option is convenient  for search 
a  vertex in large Lagrangian tables. For example, to find the 
{\it neutralino - chargino - W } vertex  one has to specify the pattens as
\verb|"~o1,~1+,W-"| . The order of patterns does not mater.

b) Automatic increase of  size of some fields in case of long input. 
This option is needed in batch sessions when program fills tables 
in blind regime and doesn't see field boundaries.

c) There is a possibility to move cursor on needed position on the 
table. This option needs to locate  position  in the table  according to 
error messages. 

d) There is an option to write the contents of given  field into 
the file and  read it back. It can be used for implementation of 
cumbersome  vertex of interaction. When the program writes to the file it 
automatically splits the field on short lines with about 80 symbols in 
each line. But when file is read, the end-of-line symbols are ignored.  

\paragraph{ 'Find' option for CalcHEP menus} is implemented. Just press
\verb|'F'| when you are in some menu and specify the needed pattern.
This option is convenient for search a record in the long menus like 
menu of SUSY 'Constraints'. Also it is used in {\it batch} calculations to 
force  the program to go to the needed menu position.

\paragraph{Comment and 'force' symbols in model tables.}
In case of tables of independent and  constrained parameters we give 
the user  possibilities   to comment the given record or  to force the 
program to include the given parameter in the numerical code. The  first 
option gives a flexibility for  model modification. The second  one 
can be used if the user  wont to have approach during numerical sessions
 to some constraints which are not involved in matrix element computation\footnote{By default 
CalcHEP does not include such parameters into numerical sessions.}.
 To comment and force a variable  
the  symbols '\%' and '*' respectively should be disposed before the parameter 
name. For example,  to include the  $b\to s\gamma$ constraint 
into numerical session mark the corresponding constraint by '*' like\\
\verb'  |  *bsgnlo|bsgnlo(mssmOk) '

\paragraph{Definition of a model} of particle interaction 
is slightly improved. Now

a) For each particle we keep its code in Monte Carlo numbering
scheme\cite{ParticlesFields}. This code is used for interface with 
structure functions routines, say CTEQ,  and interface with event 
generation packages  like PYTHIA\cite{PYTHIA}. It allows to organize 
interface that is not sensitive to name of particles used in CalcHEP.

b) In previous versions  it was forbidden to use metric tensor 
$g_{\mu\nu}$ in vertexes with spinors particles. Instead one  
had to use the anti-commutator of Dirac $\gamma$-matrices. Now this 
artificial restriction is avoided. 

c) CalcHEP works only with vetrices with simple, factored,
structure of color indexes. In order to realize vertexes like the
four-gluons 
one it was proposed  to use  auxiliary  tensor fields with color indexes
and point like propagators. Now it was recognized that in general case,
say for realization of leptoquark interactions, one needs two such 
fields. One of them plays a role of constraint, when the second presents
the corresponding Lagrange multiplier. In the current  version for each color 
vector field  two tensor auxiliary fields generated automatically\footnote{
The  names of these fields are constructed from the name of the {\it mother} field by
 adding the extensions ".t" and ".T".}.

\paragraph{Physical gauge} can by switched on/off via CalcHEP menu.
In the previous versions we  included  any model twice, first time in 
the {\it t'Hooft-Feynman} gauge and second time in the 
{\it Physical } one.
Now we keep only models defined in the {\it t'Hooft-Feynman} gauge. 
But if the user sets flag \verb|Force Unit.gauge|\footnote{
See {\it menu 2} on  Fig.  \ref{s_chain}}  in position  {\it ON}, then vertexes with 
Faddev-Popov ghosts will be ignored and propagators of massive vector 
particles will be evaluated in the {\it Physical} gauge.

\paragraph{For $1\to2$ processes} we gives the user an option to use 
effective masses of $b$ and $t$ quarks, which reproduces Higgs widths
up to all known QCD loop corrections. 

\paragraph{ $v*\sigma$ plot} for  $2\to 2$ reactions is available,
where $v$ is relative velocity. 
This option simplifies the  analysis of  $v\to 0$ limit.

\paragraph{All histograms } generated during numerical session are 
stored on disk automatically in files  \verb|distr_#|, where
\verb|#| denotes the number of current Monte Carlo session. 
The \verb|show_distr| function allows one to see  all set of generated 
distributions and to extract the needed plot. The \verb|sum_distr| 
function performs  summation of produced distribution. The examples of 
usage are \\
\verb|  ../bin/show_distr distr_1|\\
\verb|  ../bin/sum_distr distr_1 distr_2 distr_3 > distr_sum|\footnote{
As it was written above the user has approach to CalcHEP executables 
via \verb|WORK/bin|. This  example demonstrates this option.}\\
The sum is performed only for distributions which are produced in sessions 
with identical outgoing particle. It can be used for sum over incoming 
partons.

\section{MSSM in CalcHEP } 
                                   \label{MSSM}      
{\small This section follows to  \cite{micrOMEGAs}}
\subsection{Les Houches Accord for MSSM.}

Originally  the MSSM  in format of CalcHEP/CompHEP notations was realized
in \cite{MSSM}. Version of this model with effective Higgs masses evaluated
by FeynHiggsFast\cite{FeynHiggs} was presented in \cite{Semenov2}. Next step in
this business was a development of {\it Les Houches Accord}
format\cite{LesHouches} for MSSM input. This format treats all masses and
mixing angles of super-particles as independent input parameters.
It is
assumed that all these masses and angles should be   evaluated by some 
external program based  of original SUSY model.
 The list of
 parameters is presented in  Table~\ref{LesHouchesPar}.   The MSSM in {\it Les
Houches} format was implemented in framework of the micrOMEGAs
project\cite{micrOMEGAs}. Current version of CalcHEP uses namely this
realization. 

\begin{table*}[htb]
\caption{\label{LesHouchesPar}
Les Houches Accord Parameters}
\vspace{.5cm}
\begin{tabular}{|l|l||l|l|}
\hline
name     & comment                &                  name          & comment                          \\
\hline
tb        & tangent beta          &             MSnl     &$\tau$-sneutrino mass\\
alpha    & Higgs $\alpha$ angle  &              MSe$\frac{L}{R}$ & masses of  left/right selectrons \\
mu       & Higgs $\mu$ parameter  &             MSm$\frac{L}{R}$ &left/right smuon masses        \\
Mh       & Mass of light Higgs   &              MSl$\frac{1}{2}$   &masses of   light/heavy $\tilde{\tau}$ \\
MH3      & Mass of CP-odd Higgs  &              MSu$\frac{L}{R}$   &  masses of left/right u-squarks  \\
MHH      & Mass of Heavy Higgs   &              MSs$\frac{L}{R}$        & masses of left/right s-squarks \\
MHc      & Mass of charged Higgs &              MSt$\frac{1}{2}$& masses of  light/heavy  t-squarks   \\
Al       & $\tilde{\tau}$ trilinear coupling  & MSd$\frac{L}{R}$& masses of  left/right d-squarks\\
Am       & $\tilde{\mu}$ trilinear coupling   & MSc$\frac{L}{R}$& masses of  left/right c-squarks   \\
Ab       & $\tilde{b}$ trilinear coupling     & MSb$\frac{1}{2}$   &masses of   light/heavy b-squarks \\
At       & $\tilde{t}$ trilinear coupling     & Zn$_{ij}$   & i=1,..,4; j=1,..,4; neutralino mixing\\
MNE$_i$  & i=1,2,3,4; neutralino masses       & Zu$_{ij}$   & i=1,2;j=1,2; chargino U mixing \\
MC$\frac{1}{2}$ & light/heavy chargino masses & Zv$_{ij}$   & i=1,2;j=1,2; chargino V mixing \\
MSG      & mass of gluino                     & Zl$_{ij}$   & i=1,2;j=1,2; $\tilde{\tau}$ mixing \\
MSne     & $e$-sneutrino mass                 & Zt$_{ij}$   & i=1,2;j=1,2; $\tilde{t}$ mixing matrix\\
MSnm     &$\mu$-sneutrino mass                & Zb$_{ij}$   & i=1,2;j=1,2; $\tilde{b}$  mixing matrix\\
\hline
\end{tabular}
\end{table*}

We have to note that MSSM with general {\it Les Houches Accord} input
brakes the $SU(2)$ gauge invariance which leads to gauge dependence of
produced results. The implementation of large loop corrections to 
Higgs masses is done in  gauge invariant  manner\cite{Semenov2}. So, the 
problem is caused by  corrections for masses of  s-particles only. In general 
the gauge dependence should be  an order of loop corrections, because it 
caused by partial implementation of them. Nevertheless,  in some special
cases, when gauge invariance is responsible for mutual cancellations of 
diagrams, its lost can be dangerous. 

One can
check the gauge dependence by comparing results produced in Unitary gauge
against results of t'Hooft-Feynman one. 
In case of calculation of relic neutralino density\cite{micrOMEGAs} in was
detected that the gauge dependence is less than 2\%, that is smaller than
expected value of loop corrections. It looks like even for LHC energies 
small breaking of gauge invariance initiated by the {\it Les Houches Accord} is
not a problem. But this point is not absolutely clear.

After each  numerical session  the file with MSSM
parameters written in  the Les Houches Accord format appears on the disk.
The file name is  \verb|slha_#.txt|, where \verb|#| denotes the session number.
It can be used for interface  with other packages,  for example, PYTHIA.

\subsection{Parameters of MSSM.} 

The CalcHEP package   contains two   versions of MSSM. They are 
identical at the level of  {\it Les Houches Accord}, but have different 
parameter  interface. In one case parameters are specified at low energies,
whereas in the second case all  soft SUSY
breaking terms are specified  at GUT (about $10^{19}$GeV) scale. In the
second case one can significantly decrease the number of parameters  
assuming some scenario of Supersymmetry breaking.  All these
parameters  in the   CalcHEP notations  are presented in the 
Table~\ref{standardMSSM}. Note that there is a small difference between
 parameter sets used at low and  high  scales.

\begin{table*}[htb]
\caption{\label{standardMSSM} Set of  MSSM parameters. Parameters used at
the GUT scale only are marked  with \#, whereas 
parameters used only at the EWSB scale are marked with~*. Index $i$ numerates 
generations.}
\vspace{.5cm}
\begin{tabular}{|l|l||l|l|}
\hline
name& comment &name& comment\\
\hline
~tb  &  tangent beta &                      ~Ml$_i$ & Left-handed slepton masses \\
~At  & $\tilde{t}$ trilinear coupling&      ~Mr$_i$ & Right-handed selectron masses \\
~Ab  & $\tilde{b}$ trilinear coupling&      ~Mq$_i$ & Left-handed squark masses \\
~Al  & $\tilde{\tau}$ trilinear coupling&   ~Mu$_i$ & Right-handed u-squark masses \\
~Am  & $\tilde{\mu}$ trilinear coupling&    ~Md$_i$ & Right-handed d-squark masses \\
~MG1 & U(1) Gaugino mass&                  *MH3 & Mass of Pseudoscalar Higgs \\
~MG2 & SU(2) Gaugino mass&                 *mu  & Higgs $\mu$ parameter\\                                             
~MG3 & SU(3) Gaugino mass &                \#MHu & Mass of first Higgs doublet\\
\#sgn(mu)& sign of $\mu$ at low scales         &\#MHd & Mass of second Higgs doublet\\
\hline
\end{tabular}

\end{table*}

The first model  is {\tt ewsbMSSM}. Masses and mixing angles are evaluated 
by the function included in {\tt Constraints } model table,

\begin{verbatim}
suspectEWSBc(smOk,tb,MG1,MG2,MG3,Am,Al,At,Ab,MH3,mu,
           Ml1,Ml3,Mr1,Mr3,Mq1,Mq3,Mu1,Mu3,Md1,Md3,LC)
\end{verbatim}

The parameters are assumed to be given at Electroweak Symmetry Breaking 
scale $Q=\sqrt{MSt1\cdot MSt2}$. 
Besides the   parameters presented at Table~\ref{standardMSSM} it contains
auxiliary parameter {\tt smOk} included for technical purposes  and parameter {\tt LC} that switches on 
$(LC > 0)$ calculation of loop correction for s-particles. This calculation
is realized by means of SuSpect package\cite{SUSPECT}. The loop corrections 
to Higgs masses  are always included  and their calculation  also is 
supported by  SuSpect. It is assumed that MSSM parameters 
describing first two generations are identical, so only parameters 
corresponding to the first and third  generations are included into the list.

The second  model, {\tt sugraMSSM}, has the GUT scale input motivated by Super-Gravity 
 SUSY breaking scenario. Here mass spectrum is calculated by the function 

\begin{verbatim}
suspectSUGRAc(smOk,tb,MG1,MG2,MG3,Al,At,Ab,sgn,Hu,Hd,Ml1,Ml3,
                  Mr1,Mr3,Mq1,Mq3,Mu1,Mu3,Md1,Md3)
\end{verbatim}

The current {\tt sugraMSSM} model files realize  the {\it minimal SUGRA}
 scenario where \\
$Al=At=Ab = a_0$\\
$MG1=MG2=MG3 = m_{1/2}$\\
$Hu=Hd=Ml1=Ml3=Mr1=Mr3=Mq1=Mq3=Mu1=Mu3=Md1=Md3 = m_0$\\
So, there are only 4 new parameters\footnote{for some reason  in CalcHEP notations 
$ m_0$ and $ m_{1/2}$ presented as {\it mZero} and  {\it mhf}
correspondingly.}, namely, $tb , a_0, m_0, m_{1/2}$ 
and  $sgn=sgn(\mu)$ which should be $\pm 1$. 
The user can easily constructs the {\it non-minimal} sugra model moving 
some parameters from the list of {\tt Constraints} to the list of 
{\tt Variables}. An analyze of relic density and other constraints if
framework of non-minimal SUGRA was done in \cite{WMAPconstr}. 

By default we use the SuSpect\cite{SUSPECT} codes 
for solution of the Renorm-Group Equations
which connect the GUT scale parameters with low energy ones.
For other possibilities see  the next section.

Both,  the solution of RGE and the calculation of corrections to Higgs masses, 
depend on QCD sector of SM, which is not known with good 
precision. The corresponding parameters are {\tt MbMb} - scale invariant b-quark mass $Mb(Mb)$;
{\tt Mtp} - pole mass of t-quark; {\tt alfSMZ} - strong coupling constant at
MZ.  From the other side actual b and t-quarks masses  are described 
by free parameters {\tt Mb} and  {\tt Mt}.   In general case, the correct  
implementation of these masses depends on the problem under consideration. Say, for treating of
Higgs decays one has to use running  masses at the corresponding scale, but 
if these quarks appear as s-channel  resonances, then the  pole masses should be
used. In the same manner, the strong coupling which appears in QCD 
vertexes  is not defined through the {\tt alfSMZ}\footnote{ $\alpha_s$
in matrix elements depends on the used parton structure function or is driven 
by  Menu 4 of Fig.\ref{n_chain}}. 


\subsection{Isajet, SoftSusy, and SPHENO in CalcHEP.}
 For the \verb|sugraMSSM| model CalcHEP supports interface with 
with all modern RGE packages SuSpect\cite{SUSPECT}, 
 Isajet\cite{ISAJET},
SoftSusy\cite{SOFTSUSY}, and SPHENO\cite{SPHENO}. SuSpect is default one.
To use another package  
one has to replace {\it suspectSUGRAc} on \\ 
{\it isajetSUGRAc},  {\it sphenoSUGRAc}, or  {\it softSusySUGRAc}
respectively.  These functions already presented in the {\tt Constraints}
table but are commented. 

For the \verb|ewsbMSSM| model we support SuSpect and Isajet. 
In order to use Isajet for calculations masses of Higgs and super particles 
replace the \verb|suspectEWSBc| presented in list of constraints on 
\verb|isajetEWSBc| one. This function also is presented  and commented.

After model modification  one has to check installation 
of   the corresponding package and  to organize its link with CalcHEP.
In case of Isajet the  library {\it libilsajet.a} should be passed 
to CalcHEP linker. See Section \ref{UserCode}.

Interface with Spheno and SoftSusy is realized via independent call 
of the corresponding program. Parameters are passed via files written 
in framework of Les Houches Accord. The destimation of the package
is detected in runtime via environment variables {\it SPHENO/SOFTSUSY}. 
These variables  have to be defined by the user in a proper way. 
Note, that definition  depends slitely on the type of your command interpreter. 
 For example, 
if you have SoftSusy installed in subdirectory {\tt softsusy\_1.9} 
of your home directory then  define 
\begin{verbatim}
#     for sh, bash, ..           for csh, tcsh, .. 
SOFTSUSY=$HOME/softsusy_1.9    setenv SOFTSUSY ~/softsusy_1.9
export SOFTSUSY 
\end{verbatim}  

One can add  this instruction  into statup file of your Unix session
(like {\it .bashrc} or {\it .cshrc} ),
or include it in the CalcHEP statrup file \verb|WORK/calchep|. In the last 
case it should be done in the {\it sh} format and will work only 
 when  numerical  session is launched  under the symbolic one.

\subsection{MSSM constraints.}

There are several experimental constraints which allow to exclude 
some regions of MSSM  parameter space. The  references on experimental
 results, discussion  of theoretical uncertainties and details of realization
of corresponding functions see in \cite{WMAPconstr},  \cite{micrOMEGAs}.
The numbers presented below follow to  \cite{WMAPconstr}.

There is a LEP2 limit on the mass of light 
Higgs. To check it the user can control the {\tt Mh}  constrained parameter
which presents this mass.  
In case of heavy pseudoscalar Higgs  {\tt Mh>114.4 GeV}. In general case 
taking into account theoretical  and experimental uncertainties  we  have 
{\tt Mh>111GeV}. 

 To check  the LEP limits for  other  MSSM
masses  see the {\tt LEPlim} constrained parameter.  {\tt LEPlim=0}
means that all constraints are satisfied. In general case 
$$ LEPlim=l(\chi^+_1)+2l(\tilde{\nu_e} )+4l(\tilde{\nu_\mu})
+8l(\tilde{\nu_\tau}) +16l(\tilde{e_R}) +32l(\tilde{\mu_R}) +64l(\tilde{\tau_1})  \;\;,$$
where function $l()$ returns $1$ if the mass limit for the corresponding particle is 
broken are $0$  otherwise.

The {\tt drho} variable presents the $\Delta\rho$ value which describes 
the MSSM corrections to electroweak observables. It contains stop/sbotom
 contributions, as well as the two-loop QCD corrections due to 
gluon exchange and the corrections due to gluino exchange in the heavy 
gluino limit.   Precise measurements 
 allow to set  $\Delta\rho< 2\cdot 10^{-3} $.

{\tt gmuon} presents value of supersymmetric contribution to anomalous 
magnetic moment of muon. At $3\sigma$ level  
$5.1\cdot 10^{-10}<${\tt gmuon}$<64.1\cdot 10^{-10}$

{\tt bsgnlo} returns the  branching ratio for $b \to s\gamma$. 
See details of calculation in \cite{micrOMEGAs}.  Taking into 
account noticeable theoretical uncertainty of SM contribution one can set 
limits  $2.25\cdot 10^{-4} <${\tt bsgnlo}$<4.43\cdot 10^{-4} $ 

{\tt bsmumu} presents branching ratio $B_s \to \mu^{+} \mu^{-}$.
According to CDF measurements {\tt bsmumu}$<9\cdot 10^{-6}$.
The SM contribution is in 3 order of magnitude smaller.
 
These constrains were initially written for the micrOMEGAs package.
See  details in \cite{micrOMEGAs} and \cite{WMAPconstr}. It was 
demonstrated that the most strong  constraint on MSSM parameters 
follows from astrophysics  measurements of Dark Matter if it is treated 
as a relict density of neutralinos. For relic density calculations
see  the micrOMEGAs package based on matrix elements generated by 
CalcHEP.
 
\subsection{ Problem of widths for Higgs and s-patricles.}
In the current realization of MSSM the  widths of Higgs and s-particles 
are presented as free parameters. In principle they can be easily 
calculated  in a parallel CalcHEP session. The code of such auxilary session 
can be attached to the main code of cross section calculation in spirit of 
Section \ref{exportCode}. This idea 
was realized in the version of  CalcHEP included in  the
micrOMEGAs\cite{micrOMEGAs} package. But such   option is not 
realized  automatically for any  model yet. 

In the current version we pass to the user  the widths 
calculated by Isajet or Spheno. To realize  such option one 
has to  comment width variables included in the list of
 free model parameters and uncomment them in the list of constraned
parameters.  Of couse, one has to replace default SuSpect code 
used for RGE solution and loop correction on  Isajet or Spheno one\footnote
{For the \verb|ewsbMSSM| Spheno is not supported. }.

\section{QCD parton distributions}  \label{QCDPDF}     
Below we  describe two accesses
 to QCD parton distributions  supported in CalcHEP. The first of them is 
a link  of  CERN  PDF library. The second one is based on the 
special format of Particle Distribution Tables elaborated in the CalcHEP 
project. The second way allows to implement easily all  modern  CTEQ/MSRT 
parton  distributions. In both cases, by
default,  the  $\alpha_{s}$ coupling that  was used 
in  the parton distribution is applied  to matrix 
elements\footnote{ But there is a possibility to specify $\alpha_{s}$
independently.}.
For both realizations of parton distributions the user has to specify 
only incoming particle(proton/anti-proton) and identifier of the set. 
The sort of parton is detected automatically according to the Monte-Carlo
numeration defined in  CalcHEP particle tables. We also support 
special $d'(x)$,  $s'(x)$ distributions that describe quark 
densities in the models with diagonal CKM matrix.  See Section
(\ref{d_s_prim}).

 \subsection{ PDFLIB distributions}\label{PDFLIB}    

In order to include the  PDFLIB\cite{PDFLIB} distributions into the CalcHEP list
one has to pass the corresponding libraries  to CalcHEP linker. 
Then  the distributions  of PDFLIB will automatically 
appear in  CalcHEP numerical session compiled after such modification. 
While PDFLIB is not 
passes to linker, CalcHEP uses {\it dummy} version of this library.
This trick allows to install CalcHEP on computer where CERNLIB is not 
available.

The technical aspects of  attaching of  new codes to CalcHEP 
numerical sessions  are explained in 
Section \ref{UserCode}.  Note, that one can attach any QCD parton 
distributions to CalcHEP numerical session  just presenting its code 
in PDFLIB format and passing them to the  CalcHEP linker instead of 
PDFLIB.

  PDFLIB distributions for parton in nuclei are not supported  yet.

 \subsection{CTEQ and MRST parton distributions}
                                   \label{CTEQ_MRST}   

Both   CTEQ and MRST groups  store information about parton 
distributions in two-dimensional  tables and interpolate these tables.
CalcHEP has  its own file  format  for parton tables  but  uses
interpolation procedures  of CTEQ and MRST.  Thus CalcHEP produces exactly
  the same results as original CTEQ/MRST functions. 
 The information about  interpolation procedure is stored  in  CalcHEP
tables and is detected automatically.

Besides of parton distributions  CalcHEP  tables contain data for 
 $\alpha_s(q)$  which correspond to the given parton set and this function
is used by default in martix elements.

The files containing parton distributions  must  have  the  {\it "pdt"} extension. 
\verb|n_calchep|
searches such files in the directories \verb|"$CALCHEP/pdTables"|, "../",
and  "./". Usually the  last two directories are  the user's working directory and its 
sub-directory \verb|results|.

 \verb|"$CALCHEP/pdTables"| contains the following parton sets: CTEQ5M,
CTEQ6L, CTEQ6M \cite{Pumplin:2002vw} and mrstlo2002, mrst2002nlo,
 \cite{Martin:2002aw}

We pass to the user the routines  which  transform CTEQ and MRST data files
into the CalcHEP format.  By means of them the user can  add  other distribution
to the   list. The sources of these routines  are stored in the
\verb|$CALCHEP/utile| directory.
In case of CTEQ the compilation instruction is \\
\hspace*{2cm} \verb|cc  -o cteq2pdt  cteq2pdt.c alpha.c -lm |\\
For compilation one also needs the {\it alpha.h} file disposed in the same 
directory \verb|utile|.  The usage is \\
\hspace*{2cm} \verb| ./cteq2pdt < cteq_file.tbl >calchep_file.pdt|\\
for example \\
\hspace*{2cm} \verb| ./cteq2pdt < cteq6m.tbl >cteq6m.pdt|\\
The name of {\it pdt} file doesn't play a role. The \verb|cteq2pdt| routine
can be applied to any CTEQ4, CTEQ5, CTEQ6 file. It automatically detects
version of structure function and $\alpha_s$ formula.

In the case of MRST files the corresponding compilation instruction is \\
 \verb|cc  -o mrst2pdt  msrt2pdt.c alpha.c -lm |\\
The usage is \\
\verb| ./mrst2pdt |{\it name} \verb|< mrst_file.dat >calchep_file.pdt|\\
or\\
\verb| ./mrst2pdt |{\it name nf order } $\alpha(MZ)$ \verb|< mrst_file.dat > calchep_file.pdt|\\
For example \\
\verb| ./mrst2pdt |{\it mrst2002nlo 5 nlo 0.1197 }  \verb|<mrst2002nlo.dat>mrstnlo.pdt| \\ 
The number of parameters is increased in comparing with  CTEQ case, because
MRST tables don't contain the complite information. Note that {\it
name} is the  identifier  of distribution that you will see in CalcHEP
menu. If the last three parameters are not specified then  $\alpha_s$ 
will not be included in the table. See MRST documentation to find the 
proper  parameters\footnote{nf=5 always, the order is included in the file
name.} 

For  simple checks of  {\it pdt}  files one can use the {\it checkpdt} program.
The source of this program is stored in \verb|$CALCHEP/utile|. Compilation
instruction is 
\begin{verbatim}
   cc -o checkpdt checkpdt.c -I$CALCHEP/c_source/num/include \
         $CALCHEP/c_source/num/pdt.c -lm
\end{verbatim}
The usage\\
\hspace*{2cm} \verb|./checkpdt| {\it file.pdt parton x q}\\
where {\it parton} is a  parton  number according to  Monte Carlo numbering scheme. This
program  writes on the screen the corresponding parton density and
$\alpha_s(q)$. See code checkpdt.c to create more extended test.

 \subsection{d'(x) and s'(x) -parton distributions.}   
                                   \label{d_s_prim}


One can expect a noticeable  reduction  of number of diagrams 
in the models where quark mixing is absent and, so, CKM matrix is 
diagonal. Note that  the mixing between the first two generations and 
the third one is  characterized by value  about 0.04 which 
has appeared in cross sections in power 2. So, it 
can be omitted until this mixing itself is not a point 
of investigations.  Also for high energy 
physics one can neglect masses of quarks of first two generations.
Under this assumptions one can perform Cabibbo rotation in space 
of down quarks ${d,s} \to {d',s'}$ and get Lagrangian free of  mixing.

But the quark mixing being  removed from vertices re-appears in parton 
distributions. From mathematical point of view the parton distribution 
is a   quadratic form which is diagonal in the basis 
of quark mass eigenstates. After Cabibbo rotation  this form  contains
non-diagonal elements. 
$$
\left( \begin{array}{cc} d(x) &  0 \\ 0& s(x)\end{array}\right)  = 
\left(\begin{array}{cc} d'(x)\cos^2{\Theta_c} + s'(x)\sin^2{\Theta_c}  &
\frac{1}{2}\sin{2\Theta_c}(d'(x)-s'(x)) \\
-\frac{1}{2}\sin{2\Theta_c}(d'(x)-s'(x))
&  d'(x)\sin^2{\Theta_c} + s'(x)\cos^2{\Theta_c}  \end{array} \right)
$$   
It means that  now cross section contains  
products of   amplitude 
initiated  by $s'$  on  conjugated amplitude initiated by $d'$ quark.
Now we explain how one can bypass this inconvenience.

 For the processes without incoming $down$  (anti)quarks, 
the problem is absent because for these partons
the distribution form is still diagonal.
For the processes without incoming $up$ (anti)quarks
we can solve the problem  applying Cabibbo rotation for 
$up$ quarks. Thus,  actually
$$ d'(x)=d(x), \;\; s'(x)=s(x) $$
   In both these cases we have matrix elements
without mixing and with standard structure functions. 

At last we have processes  with both   $up$  and one $down$ quarks 
in initial  state.  Here  more fine treatment is needed. 
At formal level one can perform 
again Cabibbo rotation, say, for $down$ quarks and get matrix element 
without mixing in vertices. But in this particular case 
regarding the   squared diagrams one can note that the squared matrix elements
corresponding to non-diagonal reactions  are zero because of absence of
mixing in vertexes and in the $up$ quarks parton structure functions.
 So, only diagonal elements of parton density
matrix will contribute and they can be described by effective densities 
$$d'(x)=d(x)cos^2{\Theta_c} + s(x)\sin^2{\Theta_c} $$
$$s'(x)=d(x)\sin^2{\Theta_c} + s(x)\cos^2{\Theta_c}$$

CalcHEP uses  'free' Monte-Carlo
codes 81 and 83  for  $d'$ and $s'$ quarks correspondingly.
In case of CERN PDFLIB and CalcHEP PDT  distribution  
the $d'(x)$ and $s'(x)$ function are evaluated automatically
according to   $d(x)$,  $s(x)$  sorts of both
incoming partons.  

In the current version of CalcHEP package  this  $d'(x)$ and $s'(x)$
quarks are used in the SUSY models. A version of the Standard Model
with diagonal CKM matrix is stored in directory \verb|$CALCHEP/models+|.

The similar technique  based on flavor summation was presented in
\cite{FLAVORSUM}. It allows even more  economical presentation of
squared matrix elements, but used non-factored representation of 
parton distributions.

\section{ Generation of events and interface with PYTHIA.}
 \subsection{The algorithm.}       \label{vonNeumann} 

CalcHEP generates events according to the  Von Neumann 
algorithm. See   \cite{ParticlesFields}, p.202. Let the probability density 
 $f(x)$ is smaller than some  easily generated function 
$F(x)$\footnote{We assume  $f(x)$ and $F(x)$ are not normalized.}.
Then one can generate $x$ according to  distribution $F(x)$
and accept this event with probability $f(x)/F(x)$. This procedure 
is repeated in cycle until the needed  number of events is generated.

To  built $F(x)$  CalcHEP divides the space volume on 
large number of sub-cubes and in each sub-cube  sets  $F(x)$  a constant 
which equals  to $\max{f(x)}$. CalcHEP has two strategies of detecting the  
corresponding  maxima. First one is a random search. The program  generates random 
points in each 
sub-cube and tests $f(x)$ in these points. 
 The second one is a  search by the  
simplex method \cite{NumRec}.  Preliminary 
random search  is desirable  to define a  good  start point for the
simplex method.  The number of calls for random search and the number of 
steps for simplex search are  defined by the  user.

In general, the detected maxima are lower than the true ones. 
To satisfy the  inequality 
$$ f(x)\le F(x) $$
the function $F(x)$ based on the detected maxima can be 
multiplied by some factor, say $2$.  Of course, it decreases the efficiency 
of  the generator just on the same factor. Nevertheless, in some sub-cubes 
were the variance of the function is large this factor may be not enough. 
If CalcHEP finds a point $x$ where $f(x) > F(x)$ it accompany point with an
 integer weight $w$. This weight is  the integral part of
$f(x)/F(x)$   plus one with the probability 
equal to  the fraction part of $f(x)/F(x)$.  From view point of calculation of 
various  distributions  one event with integer weight $w$ should be treated as 
$w$ independent events with identical parameters. But for the evaluation 
of statistical uncertainties a more careful treatment is needed.

 \subsection{Work of generator}    \label{events}

The procedure of event generation consists of two steps. The 
first step  is a preparation (initialization)  of generator 
({\it Menu 7}). The second step is itself a generation of events ({\it Menu 8}).
 Note that the efficiency of generator should be  better if 
the user preliminary  launches VEGAS  to construct an appropriate grid  on the 
phase space.  
  
\paragraph{ Preparation of generator.}
On this step 
we divide the integration volume on several sub-cubes and find 
the  maximum of the distribution function in each sub-cube. The user has to 
define  the number of {\it Sub-cubes}, numbers of the calls of the function 
for the {\it Random~search} of maximum  and number of steps 
for the search by the {\it Simplex} method.

The larger number of  {\it Sub-cubes}, the more efficiency of the generator,
but the more time for preparation of generator is needed. The preliminary 
Vegas run   provides the user an  estimation of the  needed  time.

The preparation of generator is finished by the message which gives 
estimation of efficiency of the generator prepared.
 
\paragraph{ Work  of generator.}

Before launching the generator the user has to specify the needed
{\it Number~of~events}. In order to reduce the number of multiple 
events one can multiply the detected maxima by some factor. It should be 
done by means the second function of {\it Menu~8}. If in spite of  
it CalcHEP meets a point $x$ where the distribution function  exceeds 
the estimated maximum, them the multiple event is generated. 
 In this point the program can look for a new maximum by means of
the simplex method starting from the point $x$. The number of steps is 
defined by the third menu function. One can set this number zero to prevent
this search.

When the needed number of events  are generated, CalcHEP writes on the screen 
the message which informs the user about the efficiency of the generator 
and the number of multiple events. The user can accept the generated 
set of events or refuse them.

The generator improves itself during each run by means of the improving
the   maxima  estimation in sub-cubes. If nevertheless generator works 
badly, the user has to return to {\it Menu~7}.

The generated events are written in the text format into the file
{\it results/events\#.txt} where \# means the session number. 
Format of this file is described  in the Manual.

 \subsection{Usage of events, interface with PYTHIA.} 

We present simple routine \verb|events2tab| which constructs different 
distribution based on generated events. Format of the call is\\
\verb|../bin/events2tab|{\it Variable Min Max Nbin } \verb|<events.txt > tab.txt|\\
Here {\it Variable }  is the  name of parameter  which distribution
you would like to get. The permitted  names for {\it Variable}  
are described in Manuals.  {\it Min, Max} present boundaries of distribution, 
and $ Nbin \le 300$ is the number of bins.

We also present  routines which support  interface with PYTHIA.
The interface is done according to \cite{Belyaev-interface} and 
\cite{Boos-interface}. 

The  interface programs  are written in C and stored in the file \\
\verb|   $CALCHEP/utile/event2pyth.c|\\
We suppose that the {\it main} file will be written in Fortran.
An example is presented in \\
\verb|   $CALCHEP/utile/callPYTH.f|\\
The main program must be started from initialization routine
\begin{verbatim}
      NEVMAX= initEvents(eventFile) 
\end{verbatim} 
which opens the {\it eventFile} generated by CalcHEP and reads its header.
The return value is the number of events stored in the  {\it eventFile}.
Note that one has to remove dummy \verb|SUBROUTINE UPINIT|\\
and \verb|SUBROUTINE UPEVNT| from the PYTHIA code before link with 
CalcHEP interface program. Working versions of these routines  are disposed
in {\it callPYTH.f} and {\it event2pyth.c} respectively. 

 In case of  MSSM process one has to pass 
the corresponding models parameters  to PYTHIA. For PYTHIA 6.3 it 
can be done in framework of Les Houches Accord. Our {\it main }
\verb|callPYTH.f| contains an example of realization of such interface.
Indeed version 6.3 now is under developing and the  current version 
6.2 has not such interface. One can 'improve' PYTHIA 6.2 to solve this 
problem. Namely code for reading the Les Houches Accord file 
(written by Peter Skands for version 6.3) is disposed in \\
\verb|   $CALCHEP/utile/pyslha.f|\\
To activate this routines one has to add the corresponding call 
to PYTHIA code. In current version  \verb|pythia6225.f|
one has to  modify \verb| SUBROUTINE PYMSIN|,   inserting  after 
label \verb|120| the following code
\begin{verbatim}                                                                   
C...Read spectrum from SLHA file.
      IF (IMSSM.EQ.11.AND.IMSS(21).NE.0) THEN
        CALL PYSLHA(1,0,IFAIL)
      ENDIF
\end{verbatim}
 
The executable \verb|a.out| presenting CalcHEP-PYTHIA\_6.2  interface can be 
compiled, for instance, by 
\begin{verbatim}
   cc -c  event2pyth.c
   f77 callPYTH.f event2pyth.o  pyslha.f pythia6225.f
\end{verbatim}
And for the CalcHEP-PYTHIA\_6.3  case 
\begin{verbatim}
   cc -c  event2pyth.c
   f77 callPYTH.f event2pyth.o  pythia6225.f
\end{verbatim}

\section{Non-interactive sessions}\label{blind}   

\subsection{General concept}

CalcHEP was created for calculations in  
the interactive regime. 
 But  it is also important to have an option
 to perform   long-time calculations 
in the non-interactive, batch,  mode. Also there are  a lot of  requests for
organization of various cycles of numerical calculation when  user's
control on each step is not needed.
 It is a challenge  to support  all such kind of needs
     in the interactive package   and create  
     a comfortable  service for it.
     This is realised in the  {\it batch} session~\cite{batch}.
Instead  of  keyboard,  the program reads signals that 
simulates the keyboard hits. 
This idea  can be applied to any interactive program 
driven by the keyboard. In general, the  realization of this idea is quite  
easy. It's enough  to lock all routines that write on the screen and 
 modify one routine which   reads the keyboard signals.
It is assumed that anyway all results of calculations are stored in some
output files and, thus, they are  accessible in the same manner  after both  
interactive and batch sessions.


This way is not free of   problems. Let us list some of them:\\
a) Sometime interactive program contains branching and chooses the way by
the dialog with the user.  For example,   CalcHEP writes {\it informative} 
and  {\it dialog} messages. The first one waits  
for "Press any key", the second one expects {\it Yes/No} answer on some
question to branch the program execution.  In the batch mode  all {\it informative} 
messages are ignored, the {\it dialog} ones get the answer {\it Yes}
automatically. So, these {\it dialog} messages  should be organized in a proper 
way to support the main stream of operation.\\
b) The entry point of the program can depend on the previous session.
In this case blind simulation of keyboard can be crazy. 
We meet this problem in symbolic part of the package. See below.\\     
c) Wrong input can't be fixed in batch mode. So, the program must be 
terminated with some error code that  informs the user about problems.\\
d) Some menus  depend on the the physical problem under the consideration.
For example, one doesn't  know  a'priori the position of $t$-quark mass 
in model parameters menu.  This problem  is solved by means of the {\it
Find} menu facility which gives us a possibility to find menu record 
depending on its text label.

Thus, some improvements should  be done in  interactive programs in order to 
use them in batch mode. Usually these changes also  are welcome for the interactive
mode.

 In the CalcHEP case  the sequence of signals  is passed to the  program 
as a parameter. A special parameters, {\tt -blind} has to  precede it as  a
flag of batch regime. Thus, the batch call is realized like \\
\verb|  s_calchep -blind "STRING"|\\
\verb|  n_calchep -blind "STRING"|

Surely, the task of generating of  correct STRING  looks like an interactive 
session with closed eyes. But in the CalcHEP root directory the user 
can find  several Unix scripts which contains inside blind calls for 
most typical tasks. Parameters of these scripts are substituted  into 
the preliminary prepared sequences of commands, that allows to adapt the 
program to  user requests.

In principle,  the whole scope of problems can be solved in the batch mode.
Also   the user can  preform all needed setting in 
the interactive mode and after that launch  the batch one. This way  combines  
the  advantages of screen and batch modes.

Below we present the  CalcHEP batch command  available  and explain how
the user can writes new ones.

\subsection{Batches for symbolic calculations}
\label{s_blind}
The command 

$\bullet$ {\tt s\_blind} {\it  nModel Process nOutput}\\
performs symbolic calculation for {\it Process} in framework
of the model {\it nModel} and writes down  results in the format according to 
the {\it nOutput} parameter.
  {\it nModel} and {\it nOutput} are numbers which specify the chosen
models and output format according to items of  Menu~1 and Menu~7  of
Fig.\ref{s_chain} respectively. 
 {\it Process} must  be  enclosed into the quotation marks. For example,\\
\hspace*{1cm} \verb| ./bin/s_blind 1 "e,E->m,M" 1|\\
performs symbolic calculation in the framework of the   Standard Model 
and writes results in the C notations.  
  {\tt s\_blind} has to  be launched from user {\it WORK} directory like 
{\it calchep}.  

Because \verb|s_calchep| has different entry points {\tt s\_blind} 
preliminary removes all \verb|tmp/*| and  \verb|results/*|  files to start 
from the  beginning. 

The program of this kind is used in \cite{Belanger:2001fz} 
for runtime generation of  numerical code for various  matrix elements.

\subsection{Batches for numerical calculations}

The programs presented here are disposed in the \verb|$CALCHEP\bin| subdirectory
which in its turn is linked to user working directory.
They  should be launched from the user's subdirectory   {\it results}
like \verb|n_calchep|. So the format of call is \\
\verb|   ../bin/<batch> <Parameters>|\\
Being launched without parameters they inform  the  user about needed 
ones.

$\bullet$ {\tt run\_vegas} {\it it1 N1 it2 N2 }\\
 launches the  Monte Carlo Vegas session. In general case it launches {\tt it1}
Vegas sessions with {\it N1} integrand calls for each session, 
after that it initiates the "Clear statistics" menu function which forces the 
program to forget the obtained results and 
launches  {\tt it2}
Vegas sessions with {\it N2} integrand calls. Usually two loops are desirable 
because  in the beginning the integration grid is not adapted  to the
integrand jet. 
If {\tt it1=0} or {\tt N1=0}, only the second Vegas loop is launched
removing the   previous  results.
If {\tt it2=0} or {\tt N2=0}, only the first Vegas loop is launched
with keeping  the   previous  results.

$\bullet$ {\tt set\_momenta  p1 p2 } \\
changes momenta of incoming particles. This 
function needs two arguments,  the  values of incoming momenta.

$\bullet$ {\tt set\_param} {\it name value [... name value]}

$\bullet$ {\tt set\_param}  {\it File}\\
changes  numerical values of variables. There are two forms of its usage. 
First form is self-explanatory. In the second case it is assumed that 
{\it File}  contains two  columns, namely,  {\it name} and  {\it value}.

$\bullet$ {\tt pcm\_cycle} {\it pcm0 step N  it1 N1 it2 N2  }\\
organizes a cycle for calculation of total cross section for different 
values of particles momenta in the center-of-mass frame. {\it pcm0 }
is the initial value of momentum, {\it step } is a step of the scanning
and {\it N} is a number of steps. The {\tt  it1 N1 it2 N2}  parameters 
are passed to \verb|run_vegas|. 
 Result of calculation is stored in 
{\it pcm\_tab\_\#1\_\#2} file  where  \#1 and \#2 are the ordering numbers
of the first and the last numerical sessions.

$\bullet$ {\tt name\_cycle} {\it name val0 step N   it1 N1 it2 N2      }\\
organizes a cycle for calculation of total cross section for different 
values of the {\it name} parameter. {\it val0 }
is the initial value of parameter, {\it step } is a step of the scanning
and {\it N} is a number of steps. The {\tt  it1 N1 it2 N2}  parameters    
are passed to \verb|run_vegas|.
 Result of calculation is stored in 
{\it name\_tab\_\#1\_\#2} file  where  \#1 and \#2 are the ordering numbers
of the first and the last numerical sessions.

$\bullet$ {\tt subproc\_cycle} {\it it1 N1 it2 N2}\\
performs a cycle over all subprocesses generated.
 The {\tt  it1 N1 it2 N2}  parameters
are passed to \verb|run_vegas|.

 It is assumed that  all subprocesses have identical sets  of outgoing 
particles. In this case  one can choose  the same 
reasonable cuts, regularization and histograms for all subprocesses.  
The total cross section is displayed on the screen. If histograms are 
specified, then their  sums  will be stored in {\it dist\_\#1\_\#2} file.
The user can display them on the screen by the  \verb|disp_dist| command
like all other distributions.

\begin{figure}[ht]            

\begin{picture}(450,350)
\put(0,0){\epsfig{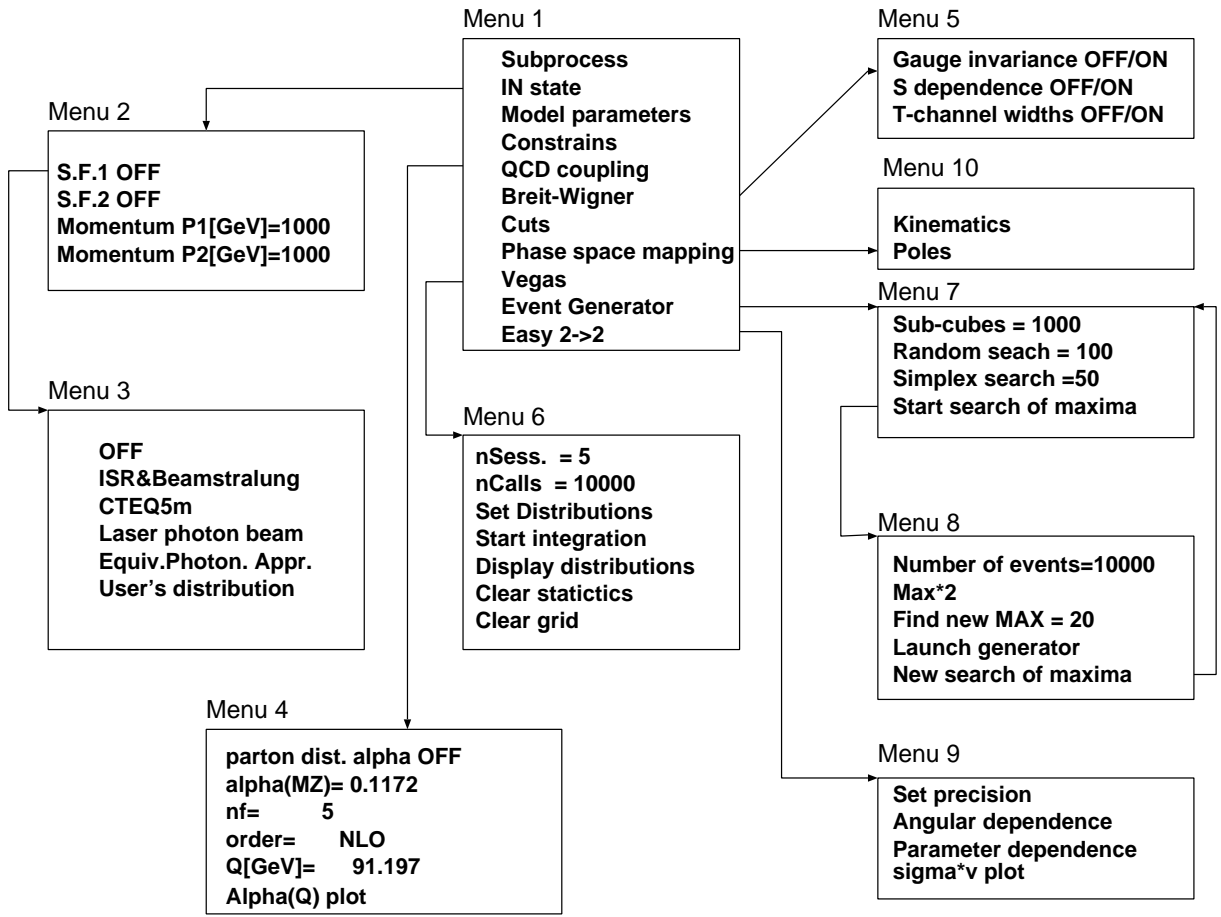}}
\end{picture}

\caption{ Menu scheme for the  numerical session }

\label{n_chain}

\end{figure}

$\bullet$ {\tt prep\_gen} {\it sub\_cubes Random\_search Simplex\_search }\\
prepares event generator. The parameters correspond to
menu~7 Fig.\ref{n_chain}.

$\bullet$ {\tt gen\_events}  {\it N\_events new\_max   } \\
generates {\it N} events. Result is written  in {\it events\_\#} file. 
The parameters correspond to menu~8 Fig.\ref{n_chain}. 

In case of error in the execution the error code  can be displayed by\\
\verb| echo $? |
The  list of possible error  codes is presented in the CalcHEP manual.

\subsection{How to write new batch.}

There is a tool for writing the command symbols  for the 
{\tt "-blind"} mode. If one launches \verb|s(n)_calchep| with the 
\verb|+blind| flag, then the program  works in the interactive
mode, but simulating all features of batch mode . After the  end of the
session the appropriated  line of commands will be   written 
on the screen. For example, let one  start\\
\hspace*{2cm} \verb|n_calchep +blind|,\\
go to Vegas menu,  set number of vegas iteration 3, specify  number of calls for 
each iteration 4444, start Vegas, and finish the session. The the line obtained  
should be:\\
\hspace*{2cm} \verb|"[[[[[[[[{{3{[{4444{[[{0"    |,\\
were '3' and '4444'  fragments corresponding to  Vegas setting. 
One can modify it, a little, to get an 
universal  Unix command for Vegas launching with parameters defined as
arguments of the command:\\
\hspace*{2cm} \verb| n_calchep -blind "[[[[[[[[{{S1{[{S2{[[{0"   ,| \\
It just one of the pieces of \verb|run_vegas| presented above.

The cording of  keyboard signals is organized by the following way.
All alphabetic character and digits  are written  naturally.
The special keys, in general, are  coded by 2-positions  hexadecimal
numbers with the  preceding   symbol "\verb|\|". For example, the 'Tab' key 
is coded  as "\verb|\08|".
In order to simplify the reading and the  modification of symbol sequence
we use the following short codes for basic special keys:\\
\hspace*{2cm} '[':   Down Arrow\\ 
\hspace*{2cm} '\{':  Enter key \\
\hspace*{2cm} '\}':  Escape key\\
\hspace*{2cm} ']':   Up Arrow key

Now one can understand the example presented above. Note, the '0'
in the end of the string appeared because the session was finished 
by pressing '0' that works like F10, which in its turn calls the  'Quit'
function.

This is a way how  the user can  create various batches without an intervention
into the C programming.  Good Luck!

\section{CalcHEP as a generator of matrix elements for 
other packages}                   \label{exportCode}  

In general one can use C-codes of matrix elements generated by CalcHEP
in other programs. CalcHEP manual contains full explanation  of format 
of  generated routines. Also we present a simple example  \\
\verb|$CALCHEP/utile/main_22.c|,
which demonstrates the usage of generated codes for $2\to2$ reactions.
The compilation instruction launched from the directory \verb|results|,
where  C-codes for  some $2\to2$ process are generated  should be
\begin{verbatim}
 cc -I$CALCHEP/include $CALCHEP/utile/main_22.c \
   $CALCHEP/c_source/sqme_aux/sqme_aux.c  *.c   -lm
\end{verbatim}
 
We foresee a possibility to combine in one project several matrix elements 
produced by CalcHEP. The names of routines generated by CalcHEP are 
finished by the the postfix
\verb|"_ext"|, which can be replaced on other one to avoid name conflicts.
Also we have organized one {\it structure \tt interface\_ext}, which contains all interface routine, 
so the user can change the code for matrix element by means of one
operation.

The command\\
 \verb|  ../bin/mkLibstat |{\it postfix}\\
started from the \verb|results| directory replaces "ext" on your 
{\it postfix}, compiles all routines and transforms them into 
\verb|proclib_|{\it postfix}\verb|.a| library which is ready to by included 
into other packages. 

The similar command \\
 \verb|  ../bin/mkLibldl |{\it postfix}\\
created the  shared \verb|proclib_|{\it postfix}\verb|.so| library 
that can be loaded dynamically in run time. 

There is  a very attractive possibility to generates and link automatically 
 new matrix elements when they becomes needed. New codes 
can be generated by a command like \verb|s_blind| described in Section
\ref{s_blind},  compiled by \verb|mkLibldl| and dynamically loaded by 
the standard \verb|dlopen| routine. 

 This approach was realized in the
micrOMEGAs\cite{Belanger:2001fz,micrOMEGAs} project
were numerous   co-annihilation processes potentially needed for
calculation  of neutralino relic density are added to the package in runtime 
depending on mass relations between super particles.

\section{Loading  of additional  codes in CalcHEP.} 
                                  \label{UserCode}    
There are several cases when facilities of CalcHEP numerical session 
can be enhanced by implementation of addition  codes. They  are 
a) CERN PDFLIB parton distributions (Section  \ref{PDFLIB}),
b) user  functions for distributions and cuts,
c) constraints for  new models.

User's startup file \verb|calchep| contains definition of 
environment variable {\it EXTLIB }, whose contents is passed to the routine 
which creates \verb|n_calchep|. Default it is defined as
\begin{verbatim}
   EXTLIB="$CALCHEP/susylib.a"
\end{verbatim}
and passes to linker the library of SUSY  constraints and {\it SuSpect} RGE codes.
If the user would like to add CERN PDFLIB, he has to modify EXTLIB like 
\begin{verbatim}
EXTLIB="$CALCHEP/susylib.a -L/cern/pro/lib -lpdflib804 \
                          -lmathlib -lpacklib"
\end{verbatim}
Here we assume that CERNLIB is disposed  in its {\it standard} place.
To add {\it Isajet} RGE code disposed, say, in user's directory
\verb|isajet|, the needed modification is 
\begin{verbatim}
EXTLIB="$CALCHEP/susylib.a $HOME/isajet/libisajet.a"
\end{verbatim}
Depending on switches used for {\it Isajet} compilation, may be, one
needs to add CERN {\it mathlib} too.

If one  implements  new model in CalcHEP which has addition constraints, 
the library of these  constraints  can be attached   to \verb|n_calchep| 
by the  same manner. It should be a library of functions written in C with 
arguments and returned values of type {\it double}. 

Other useful possibility is the implementation of the \verb|usrfun| routine which should 
be a function written in C with argument of the {\it "char~*"} type
and returning {\it double}.  Then the user can construct the family of  cuts and 
distributions named like \verb|Uxxx|, where "xxx" will be passed to 
\verb|usrfun| as an identifier of the function.  In the  Manual it is explained how one 
can pass momenta of particles to such function.   

We have to note that anyway CalcHEP passed to linker dummy 
libraries for PDFLIB and \verb|usrfun()| function.  \verb|$EXTLIB|
is substituted in list of abjects for linking before these dummy 
libraries. So \verb|n_calchep| will be linked successfully, does not 
matter it gets or not  the true objects.

\section {Contributions and References}

In general, references depend on which path of the package is used.

The kernel of the package was written  by Alexander Pukhov, cite
\begin{verbatim}
   A.Pukhov et al, Preprint INP MSU 98-41/542,arXiv:hep-ph/9908288
   or the current paper.
\end{verbatim}
Interface with different RGE codes as well as  $(g-2)_{\mu}$, $b->s\gamma$,
and $B_s->\mu^+\mu^-$ constraints were  written by   microMEGAs team. 
If you use it, please, cite
\begin{verbatim}
  G.Belanger, F.Boudjema, A.Pukhov, A.Semenov,
  Preprint LAPTH-1044, arXiv:hep-ph/0405253.
\end{verbatim}

The default RGE code included into the package is SuSpect(version 2.3).
If you use it, please, cite  
\begin{verbatim}
  A.Djouadi, J.-L.Kneur and G.Moultaka,
  Preprint  PM-02-39, CERN-TH-2002-325, arXiv:hep-ph/0211331.
\end{verbatim}

The package contains  some codes written by CompHEP group people.
They are included in the CalcHEP package with permission of the authors: 
\begin{verbatim}
V.Ilyin    : num/4_vector.c,num/strfun/sf_epa.c,num/strfun/sf_lsr.c
D.Kovalenko: num/kininpt.c,num/regfunal.c
A.Kryukov  : symb/colorf.c,chep_crt/edittab.c
V.Edneral  : symb/diagram/diaprins.c,symb/diagram/drawdiag.c,
             chep_crt/crt.c
A.Semenov  : chep_crt/xwin/X11_crt0.c
\end{verbatim}

\section{Acknowledgments }

The work was supported by   GDRI-ACPP of CNRS and grant of Russian Federal Agency for Science NS-1685.2003.2. 
Large part of this work was motivated by development 
of microMEGAs project\cite{micrOMEGAs}. Author thanks G.Belanger,
F.Boudjema, and A.Semenov for mutual development of these codes and numerous 
discussions. I thanks also to A.Belyaev and A.Datta for
suggestions,discussions, and the testing.

\newpage
\section*{Pictures and figures}

\end{document}